\documentstyle[prl,aps,epsfig,amsfonts]{revtex}

\begin{document}

\twocolumn \psfull \draft

\wideabs{
\title{Suppression of the ``quasiclassical'' proximity gap in
correlated-metal--superconductor structures}
\author{Branislav K. Nikoli\' c,$^\dagger$ J. K. Freericks,$^\dagger$ and
P. Miller$^*$}
\address{$^\dagger$Department of Physics, Georgetown University,
Washington, DC 20057-0995 \\
$^*$Department of Physics, Brandeis University, Waltham, MA 02454}

\maketitle

\begin{abstract}
We study the energy and spatial dependence of the local density
of states in a superconductor--correlated-metal--superconductor
Josephson junction, where the correlated metal is a non-Fermi
liquid (described by the Falicov-Kimball model). Many-body correlations
are treated with dynamical mean-field theory, extended to inhomogeneous
systems. While quasiclassical theories predict a minigap in the spectrum
of a disordered Fermi liquid which is proximity-coupled within a mesoscopic
junction, we find that increasing electron correlations destroy any minigap
that might be opened in the absence of many-body correlations.
\end{abstract}

}

\narrowtext

Fermi-liquid metals have excitation spectra that typically vary on energy
scales of electron volts.  Metals that become superconducting, have all
low-energy electrons form pairs.  Since it takes an energy
equal to the superconducting gap $\Delta$ to break a pair of electrons and
form an excitation, there is a low-energy gap (on the order of meV)
in the single-particle density of states (DOS)\cite{bcs}. The original
states at the Fermi level are ``pushed'' to excitations near $\pm\Delta$, which
yields a singularity in the DOS at zero temperature (and a large peak at
finite $T$).  When a superconductor ($S$) is connected to a normal metal
($N$) to form a  $SNS$ Josephson junction, the superconductivity
leaks into the normal metal via the proximity effect\cite{degennes},
and a weak link is established between the two $S$ through the $N$.
What happens to the low-energy electrons in the $N$ is quite interesting.  An electron
near the Fermi level of the $N$ is localized within the $N$ because there
are no single-particle states at low energy for it to scatter into within the
$S$.  Instead, the electron is retroreflected into a hole in the $N$, and creates
a superconducting pair in the $S$ via a process called Andreev
reflection\cite{andreev}.  This reflection occurs at the $SN$ interface and the
$NS$ interface, creating Andreev bound states with well-defined energy levels.
These states are doubly degenerate, one carrying supercurrent to the right
and one to the left.  The states are broadened into peaks in the DOS when one
averages over all different perpendicular momenta. Thus, Andreev reflection mixes the
electron and hole states in the same proportion that they are mixed to form
Bogoliubov quasiparticles in the $S$, with weights determined by the self-consistency
condition. Such partially superconducting properties of the proximity-coupled
normal metal are responsible for Josephson effect, as well as
other peculiar phenomena in inhomogeneous systems which have
been drawing increased attention over the past decade due to advances
in mesoscopic superconductivity~\cite{lambert}.

Another aspect of the proximity effect is the modification of the
DOS in both the $S$ (``inverse proximity
effect''~\cite{inverse_exp,inverse}) and the $N$ side of a $SN$
boundary, which becomes most conspicuous in
mesoscopic~\cite{lambert} confined geometries~\cite{golubov97}.
For example, in a sufficiently long $SNS$ junction at low enough
temperature, the proximity of the superconductor induces a
minigap in the local density of states (LDOS) inside $N$
interlayer~\cite{golubov97,inverse,zhou} that has chaotic
classical dynamics~\cite{lodder,rmt}. The minigap is of the order
of the Thouless energy $E_{\rm Th}=\hbar/t_{\rm dwell}$, where
$t_{\rm dwell}$ is the typical time spent by an electron during
its diffusive motion~\cite{golubov97,zhou,belzig} ($t_{\rm dwell}
\simeq L^2/{\mathcal D}$ for a $N$ strongly coupled to a $S$, where
${\mathcal D}$ is the diffusion constant) or during its chaotic
ballistic motion~\cite{rmt} through the $N$ region of size $L$,
before escaping into the superconductor (for integrable classical
dynamics in the $N$, the LDOS is nonzero, but vanishes nearly
linearly at the Fermi level~\cite{mcmillan}). These results were
obtained using either quasiclassical
approaches~\cite{inverse,golubov97,zhou,lodder,belzig} or
mean-field treatment by random matrix theory~\cite{rmt}. However,
recent calculations~\cite{ostrovsky}, which include additional
quantum effects through a supersymmetric non-linear $\sigma$-model
(NLSM)~\cite{efetov}, show that mesoscopic fluctuations cause an
exponentially small smearing of the quasiclassical gap in a
diffusive $SNS$ junction (the DOS tails appearing below the
quasiclassical gap edge are due to prelocalized
states~\cite{efetov} which couple weakly to the $S$ leads). Using
tunneling spectroscopy, the proximity affected LDOS can be
measured as a function of the distance from the $SN$ interface in
both the normal metal~\cite{gueron} and
superconductor~\cite{inverse_exp}.

\begin{figure}
\centerline{\psfig{file=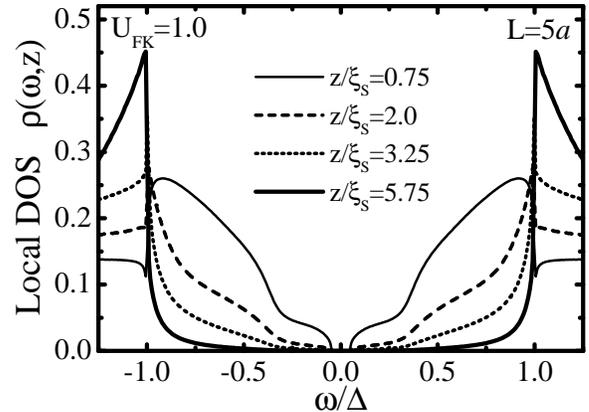,height=3.0in,angle=-90} }
\vspace{0.2in} \caption{Local DOS in the superconducting side
of the $SCmS$ junction with thickness $L=5a$ and the $Cm$
interlayer described by the FK model with $U_{\rm FK}=1.0$. The
planes are labeled by their distance from the center ($z=0$)
of the $Cm$ region, in units of the superconducting coherence
length $\xi_S \approx 4a$ (the $SCm$ interface occurs at $z=0.75 \xi_S$). The
region inside the semi-infinite $S$ leads where the self-consistent
calculation is performed extends to $z=8.25 \xi_S$.} \label{fig:superldos}
\end{figure}

While both quasiclassical and NLSM calculations rely on the
picture of well-defined (noninteracting) quasiparticles, little
attention has been paid to proximity effects in systems where the
$N$ layers are dominated by strong many-body correlations (except
in one dimension where anomalously enhanced DOS have been found
in a Luttinger liquid coupled to a superconductor~\cite{luttinger}).
Here we explore the LDOS, in both the superconducting and correlated
metal sides of a superconductor--correlated-metal--superconductor ($SCmS$)
Josephson junction. The  
\begin{figure}
\centerline{\psfig{file=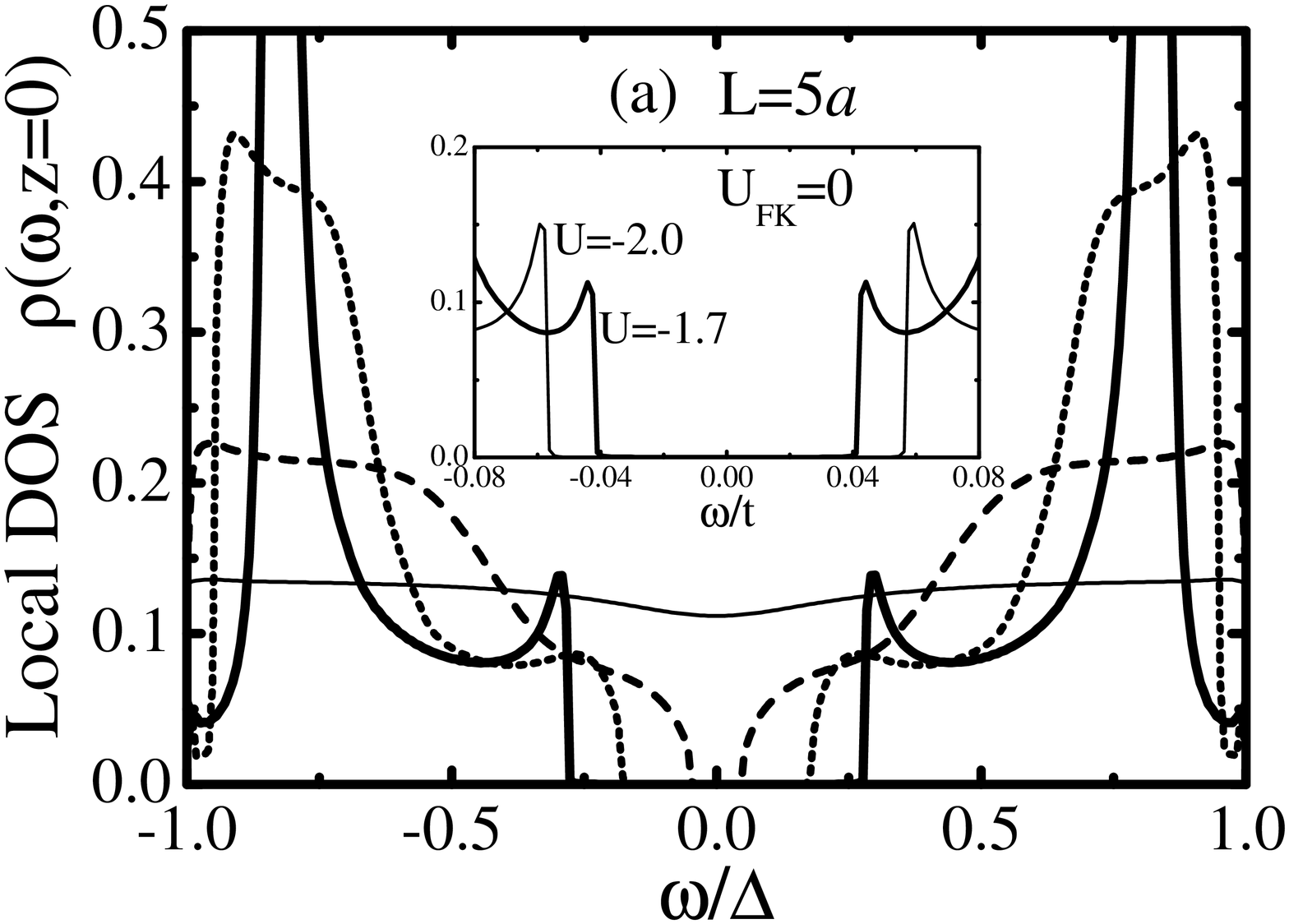,height=3.0in,angle=-90} }
\vspace{0.2in}
\centerline{\psfig{file=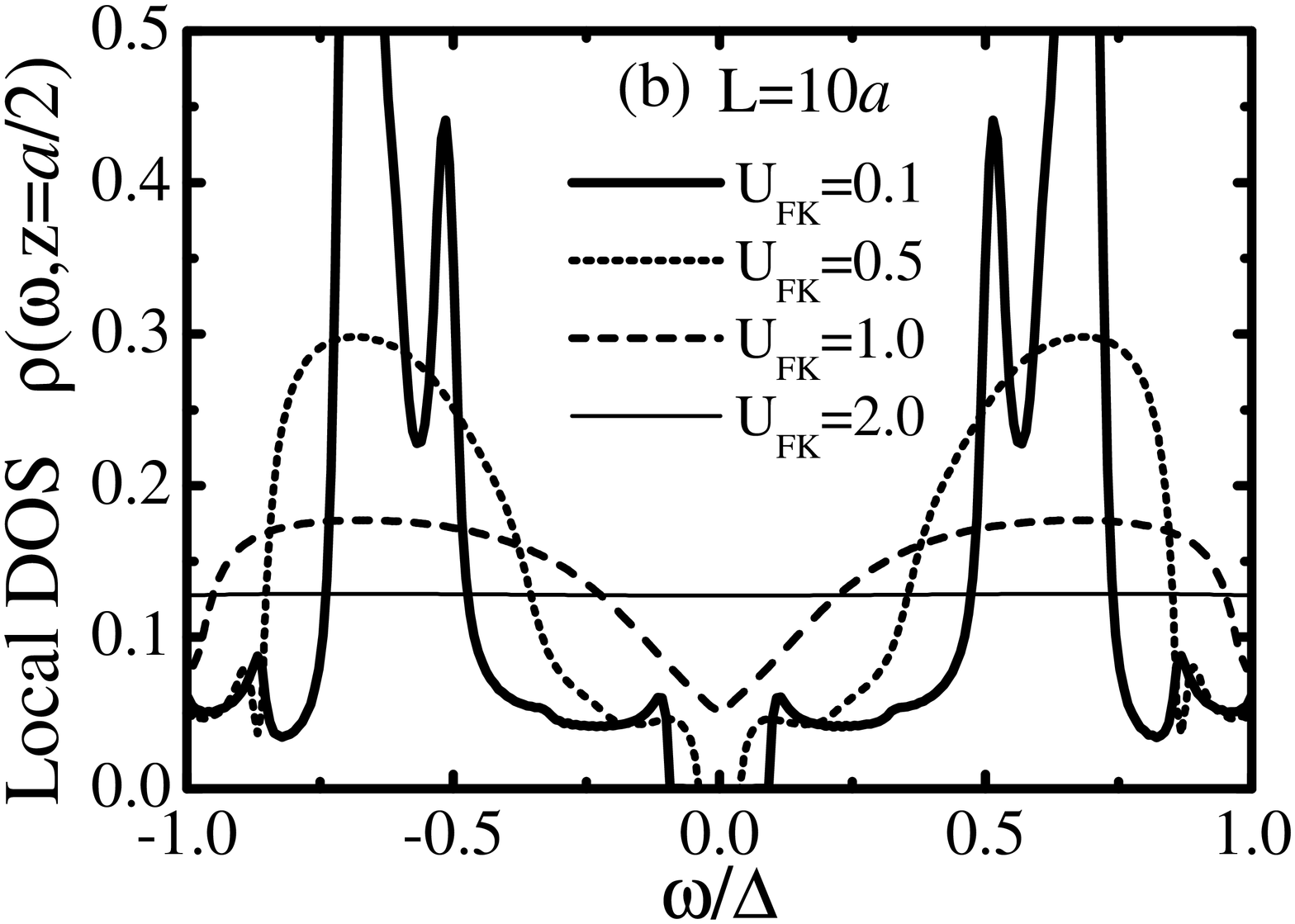,height=3.0in,angle=-90} }
\vspace{0.2in} \caption{Local DOS on the normal plane closest to
the center ($z=0$) of $SCmS$ Josephson junctions of thickness:
(a) $L=5a$ and (b) $L=10a$. The $Cm$ interlayer is a non-Fermi
liquid described by the FK model where the bulk DOS is flat in
the plotted energy range around the band center $\omega=0$ [e.g.,
$N(\omega)=0.127$ at $U_{\rm FK}=2$, see Fig.~\ref{fig:fkbulk}].
The minigap in the LDOS of the $Cm$ is open only for small enough
$U_{\rm FK}$, and does not scale as $\sim 1/L^2$ which would be
the quasiclassical prediction, but seems to be just a remnant of
the minigap $\sim \Delta^2/\mu$ [inset in panel (a), where
$\Delta(U=-2)/\Delta(U=-1.7) \approx 1.9$] in a clean $SNS$
junction which is gradually destroyed by increasing the strength
of many-body correlations. } \label{fig:minigap}
\end{figure}
 $Cm$ region is a non-Fermi liquid modeled by a
Falicov-Kimball (FK) Hamiltonian~\cite{falicov}. We find that
increasing electron correlations completely destroy any initially
open ``quasiclassical'' minigap in the DOS of the correlated
metal, as shown in Fig.~\ref{fig:minigap}; this occurs due to the
extensive broadening of the ``Andreev bound states'' by the
scattering in the $Cm$. The appearance of a nonzero LDOS inside
the $S$ region within a distance on the order of the
superconducting coherence length $\xi_S$ is plotted in
Fig.~\ref{fig:superldos}. Our analysis is fully self-consistent
(i.e., we take into account the suppression of the superconducting
order parameter inside the $S$ leads) at zero Josephson current across the
junction. The calculation includes all (many-body) quantum effects
encompassed in the 
dynamical mean-field theory~\cite{dmft}, which has only 
recently been generalized to treat inhomogeneous normal
systems~\cite{pott} and Josephson junctions~\cite{miller}.

Thus, our principal result is substantially different from the
standard lore of a proximity-induced ``hard minigap'' (i.e., no
states inside an energy interval $E_g \sim E_{\rm Th}$), which is
supported by both quasiclassical
calculations~\cite{golubov97,zhou,belzig}
and the picture of bound states induced by
Andreev reflection~\cite{andreev}. Such qualitative
considerations~\cite{belzig} give an estimate for $E_g$, which
comes close to the values of the quasiclassical minigap obtained
from the solution of the Usadel equation: $E_g=0.78E_{\rm
Th}$ in the diffusive $N$ layer of an $INS$
structure~\cite{belzig} ($I$ is an insulator that specularly
reflects), or $E_g=3.12 E_{\rm Th}$ in a diffusive $SNS$
junction~\cite{zhou}. The perturbative analysis of the
electron-electron interaction, within quasiclassics, only
generates a slightly smaller $E_g$~\cite{zhou}. The NLSM
calculations~\cite{ostrovsky} for a $SNS$ junction, with $N$ being
a disordered noninteracting electron system, find nonzero DOS at all
$E < E_g$, but the tail of subgap states is small for good metals (except 
near $E_g$). Our results are {\em complementary} to these theories, 
showing how the minigap can disappear at a critical correlation 
strength, and are of special interest in understanding the limitations 
of a phenomenological application of standard proximity-effect theory 
to experiments dealing with unconventional inhomogeneous structures, 
such as high-$T_c$ Josephson junctions with underdoped cuprates 
(a strongly correlated electron system) playing the role of the
``normal region''~\cite{htc}.

The $SCmS$ Josephson junction is modeled by a Hamiltonian
\begin{eqnarray}
H&=&-\sum_{ij\sigma}t_{ij}c_{i\sigma}^{\dag}c_{j\sigma}+\sum_iU_i\left
( c_{i\uparrow}^{\dag}c_{i\uparrow}-\frac{1}{2}\right ) \left (
c_{i\downarrow}^{\dag}c_{i\downarrow}-\frac{1}{2}\right )\cr
&+&\sum_{i\sigma}U_i^{FK}c_{i\sigma}^{\dag}c_{i\sigma}\left (
w_i-\frac{1}{2} \right ),
\end{eqnarray}
on an infinite set of stacked square lattice planes, whose connectivity
is the same as a simple cubic lattice (with lattice constant $a$).
Here $c_{i\sigma}^{\dag}$ ($c_{i\sigma}$) creates (destroys) an
electron of spin $\sigma$ at site $i$, $t_{ij}=t$ (the energy unit)
is the hopping integral between nearest neighbor sites $i$
and $j$ (both within the planes and
between planes), $U_{i}<0$ is the attractive Hubbard interaction for sites
within the superconducting planes, $U_{i}^{FK}$ is the
FK interaction for planes within the $Cm$
region, and $w_{i}$ is a classical variable that equals 1 if an
$A$ ion occupies site $i$ and is zero if a $B$ ion occupies site
$i$. The chemical potential $\mu$ is set equal to zero to yield
half filling in the $S$ and $Cm$.
The negative-$U$ Hubbard term describes the real-space
pairing of electrons due to a local instantaneous attractive
interaction. This generates a superconducting order in the $S$
leads which, when  treated in the Hartree-Fock approximation, is
equivalent to conventional BCS theory, except that here the
DOS is non-constant and provides the energy cutoff.

\begin{figure}
\centerline{\psfig{file=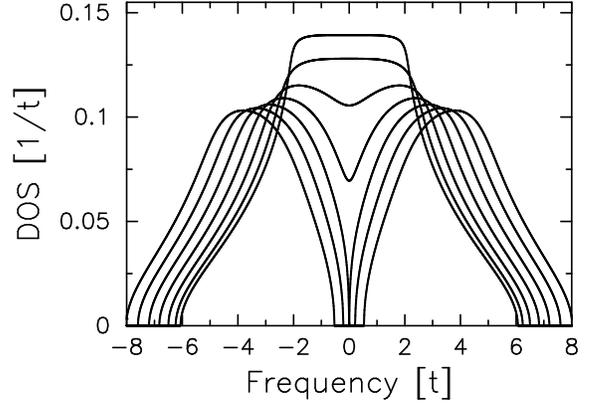,height=2.2in,angle=0} }
\vspace{0.2in} \caption{Electronic DOS (per spin) for the bulk $Cm$
described by the FK model on a simple cubic lattice
in the local approximation. The value of $U_{FK}$ ranges from 1
to 7 in steps of 1. As $U_{FK}$ increases, the DOS first develops a
pseudogap and then a real gap. We use $U_{\rm FK} \le 2$ for the
$Cm$ region of the $SCmS$ Josephson junction
(Fig.~\ref{fig:minigap}) where the bulk DOS is constant around
half-filling.} \label{fig:fkbulk}
\end{figure}

The superconducting layers have $U_{i}=-2$ and $w_{i}=0$ for all
sites. Such a homogeneous bulk superconductor is characterized by
the usual BCS parameters: the transition temperature $T_c=0.11t$,
the zero-temperature order parameter $\Delta=0.198 t$, and the coherence
length $\xi_S=\hbar v_F/(\pi\Delta) \approx 4a$. The Cm
interlayer is described by a half-filled FK model in
the symmetric limit of half filling for the ``ions''
$\langle w_{i} \rangle=0.5$.  One can view the FK
metal as a binary alloy of $A$ and $B$ ions at 50\% concentration
with $U_{FK}$ being the difference in site energy between the $A$
and $B$ ionic sites.  The many-body problem is solved by taking an
annealed average that yields the coherent potential
approximation. This is the simplest many-body problem which,
nevertheless, mimics a metal-insulator transition
(MIT) of the type seen in the repulsive Hubbard model (except the metallic phase
is not a Fermi liquid here). In the bulk, the FK correlated metal
undergoes a MIT at $U_{FK}\approx 4.9t$ (which is close to half of the
bandwidth $6t$). This is illustrated in Fig.~\ref{fig:fkbulk}
which shows the DOS in a bulk $Cm$ as a function of
$U_{FK}$. The DOS is independent of temperature~\cite{vandongen}.
Since the system is not a Fermi liquid for nonzero $U_{FK}$, the
DOS first develops a pseudogap, and then is suppressed entirely
to zero as the correlations increase, eventually driving the
system into a correlated insulator.  The opening of the gap is
continuous. In order to focus only on the modification of the DOS
induced by the proximity effect, we choose $U_{\rm FK} \le 2$ for
the strength of Coulomb interaction in the FK correlated metal,
which ensures that the bulk DOS around the band center is essentially constant.

The problem of inhomogeneous superconductivity
is solved by employing the Nambu-Gorkov matrix formulation for
Green functions with a local self-energy~\cite{miller}. We treat
the problem self-consistently in the complex order parameter
$\Delta_i$ for the part of the junction comprised of the normal
region (containing 5 or 10 planes) and 30 superconducting planes
on each side of the $Cm$ interlayer. Inside this
``self-consistent-part'' of the infinite $SCmS$ junction, the
superconducting gap $\Delta_i$ heals to its bulk value since all
signatures of the inverse proximity effect are gone on the length
scale of few $\xi_S$ away from the $SCm$ interface. The
calculation is performed at the temperature $T=0.09T_c$ where the
BCS gap is fully developed. Details of our computational
algorithm have been given elsewhere\cite{miller}. The final
result is the self-consistent Green function which allows us to
compute the (many-body) LDOS as a function of plane position
$z_i$ and frequency $\omega$ from the real-axis analytic
continuation, $\rho(\omega,z_i)=-{\rm Im}\, G(\omega+i\delta
;z_i,z_i)/\pi$.

The strength of the superconductivity in the $N$ interlayer is
quantified by the nonzero pair amplitude $F(z_i)=\Delta_i/|U_i|$
(a two-particle property) which decays exponentially, due to the
absence of an attractive interaction, on the length scale
$\xi_N=\sqrt{\hbar {\mathcal D}/2\pi k_B T}$~\cite{degennes} (or
as a power law at zero temperature  in a clean normal
metal~\cite{falk}). This is the length scale over which two
thermal electrons with energy $\omega \approx \pi k_BT$ in the
$N$, correlated  by Andreev reflection over the length scale
$L_\omega=\sqrt{\hbar {\mathcal D}/2\omega}$, lose their relative
phase coherence (which then determines the thermodynamic critical
Josephson current $I_c$). However, single-particle properties
(like the DOS)~\cite{inverse,luttinger}, or kinetic
quantities~\cite{golubov97}, can be influenced on a much longer
length scale (where Josephson coupling vanishes) $L_\omega \gg
\xi_N$ for low-energy electrons $\omega \ll kT$, which is
ultimately limited by the mesoscopic phase-breaking length
$L_\phi$~\cite{lambert}. Therefore, the anomalies in the DOS for
low $\omega$ extend up to the energy dependent distances $\sim
L_\omega$~\cite{inverse,gueron} from the $SN$ interface (which is
smeared upon approaching $L_\phi$~\cite{gueron}). In a ``closed''
geometry, where a finite-size $N$ is disconnected from any electron
reservoirs~\cite{golubov97}, this leads to a position independent
minigap edge at $E_g \sim E_{\rm Th}$ for sufficiently long
$E_{\rm Th} \ll \Delta$ diffusive interlayer ($\ell \ll L$, where
$\ell$ is the mean free path). Thus, both the quasiclassical and
NLSM descriptions of the proximity effect rely on the essential
concepts in disordered Fermi-liquid physics, such as $E_{\rm Th}$
governing thermodynamic and quantum transport phenomena in
mesoscopic systems. Since our $Cm$ layer is a non-Fermi liquid,
these concepts are not directly transferable. Therefore, to
compare our findings with standard notions, we proceed along a
phenomenological route frequently (but unwarrantedly) employed in
experiments on $SNS$ Josephson junctions with an unconventional
$N$ interlayer~\cite{htc}. Namely, we extract an effective
diffusion constant from the Kubo conductivity of the FK
model~\cite{miller}, using the Einstein relation $\sigma_{\rm
FK}=2 e^2 N(0) {\mathcal D}$ and the DOS at half-filling $N(0)$
from Fig.~\ref{fig:fkbulk}, and then compute $E_{\rm Th}=\hbar
{\mathcal D}/L^2$. It is interesting to check if such an energy
scale provides any heuristic guidance in interpreting our results
at small $U_{\rm FK}$ where the hard minigap is present in the
$Cm$ spectrum. For example, in the largest $U_{\rm FK}=2.0$
sample, the resistivity of the FK correlated metal is $\rho_{\rm
FK}\simeq 240$ $\mu \Omega$cm (assuming $a=3$ \AA). From here we
get ${\mathcal D} \approx 2 ta^2/\hbar$ and $\xi_N \simeq 5.6a$.
This is surprisingly close to the true $\xi_N \approx 6.7a$
extracted from the decay of $I_c$ in the $SCmS$ junction as a
function of the $Cm$ layer thickness~\cite{miller} (the agreement
improves for smaller $U_{FK}$). The quasiclassical analysis
for a mesoscopic diffusive junction of the same resistivity and
with thickness $L=10a$ would give $E_{\rm Th} \approx 0.1 \Delta$
and  $E_g \approx 0.32 \Delta$. However, at $U_{\rm FK}=2.0$ no
gap is found in the $Cm$ spectrum, while only a small dip
(Fig.~\ref{fig:minigap}) in the LDOS persists as a remnant of the
minigap opened for $U_{\rm FK} \lesssim 1.0$. Moreover, the
ratios of the minigap sizes $E_g^a/E_g^{b}$ in the $SCmS$
junctions with two different thicknesses $L_a=5a$ and $L_b=10a$
are: (i) $U_{\rm FK}=0.1 \Rightarrow E_g^a/E_g^{b}=2.9$, (ii)
$U_{\rm FK}=0.25 \Rightarrow E_g^a/E_g^{b}=3.1$, (iii) $U_{\rm
FK}=0.5 \Rightarrow E_g^a/E_g^{b}=4.4$, is a function of $U_{\rm
FK}$. These are different from the expected results of
$E_g^a/E_g^{b}=(L_b/L_a)^2=4$ or  $E_g^a/E_g^{b}=L_b/L_a=2$,
which  would follow from $E_g \sim E_{\rm Th}$ analogy with the
quasiclassical description of the proximity effect in a
diffusive  Fermi-liquid metal of the same
resistivity~\cite{zhou,belzig}, or a clean but chaotic
interlayer~\cite{lodder}, respectively.

The last two pieces of information needed to characterize proximity-induced 
effects in correlated metals are: 
(1) the LDOS is position dependent, while $E_g$ is spatially
constant (which is the same as the quasiclassical
phenomenology~\cite{belzig}), and does not change upon lowering
the temperature below our reference $T/T_c=0.091$ for
Fig.~\ref{fig:minigap}; (2) the minigap is open for small enough
$U_{\rm FK}$, where Andreev bound states also clearly coexist
with it, as shown by the peaks below $\Delta$ in the $U_{\rm
FK}=0.1$ case in Fig.~\ref{fig:minigap}. In fact, we find the
largest minigap in the limit $U_{\rm FK} \rightarrow 0$
(corresponding to a clean $SNS$ junction), which is generated by
the normal reflection at the $SN$ interface due to a
non-negligible $\Delta/\mu \approx 0.03$ determining the
amplitude of the scattering~\cite{hurd} ($\mu$ is the Fermi
energy measured from the bottom of the band). It appears that
increasing of electron correlations by increasing $U_{\rm FK}$ 
then just leads to a monotonic vanishing of any 
initially open minigap in the noninteracting case 
[$E_g(U_{\rm FK}=0)$ depends on $U_i$, as illustrated in the inset 
of Fig.~\ref{fig:minigap}, and is outside of the quasiclassical 
approximation~\cite{hurd}, but belongs to the realm of ``noninteracting 
quasiparticle'' physics]. This should be contrasted with the quasiclassical 
minigap as a function of quenched disorder~\cite{belzig}: the 
minigap forms for arbitrarily small concentration of impurities,
increases with $1/\ell$ to a maximum value when $\ell \sim L$,
and then decreases in the diffusive limit as $E_{\rm Th} \sim
v_F\ell$. On the superconducting side of our $SCmS$ junction we
find that position dependent LDOS is nonzero in the energy range
$2\Delta$ (Fig.~\ref{fig:superldos}), and decays to zero on the
length scale of a few $\xi_S$ from the $SCm$
boundary~\cite{inverse}.

{\it Acknowledgments:}
Support from ONR grant number N00014-99-1-0328 is
gratefully acknowledged.  Computer calculations were partially
supported by HPC time from the Arctic Region Supercomputer
Center.

\end{document}